# Structural phase transformation and phase boundary/stability studies of field-cooled Pb(Mg$_{1/3}$Nb$_{2/3}$O$_3$)-32%PbTiO$_3$ crystals

Hu Cao, Feiming Bai, Jiefang Li, and D. Viehland
*Dept. of Materials Science and Engineering, Virginia Tech, Blacksburg, VA 24061*

Guangyong Xu, H. Hiraka* and G. Shirane
*Dept. of Physics, Brookhaven National Laboratory, Upton, NY 11973*

Structural phase transformations in (001)-oriented (1-x)Pb(Mg$_{1/3}$Nb$_{2/3}$O$_3$)-32%PbTiO$_3$ (PMN-x%PT) crystals have been investigated by x-ray diffraction. A C→T→M$_C$ sequence was observed in both the field-cooled (FC) and zero-field-cooled (ZFC) conditions. Most interestingly, an anomalous increase in the C→T phase boundary with increasing field has been observed, which is seemingly a common characteristic of crystals whose compositions are in the vicinity of the MPB, irrespective of the width of the T and M$_C$ phase regions.

Single crystals of the complex perovskite systems (1-x)Pb(Mg$_{1/3}$Nb$_{2/3}$O$_3$)-xPbTiO$_3$ (PMN-x%PT) and (1-x)Pb(Zn$_{1/3}$Nb$_{2/3}$O$_3$)-xPbTiO$_3$ (PZN-x%PT) have exceptional electro-mechanical properties[1,2]. The ultrahigh piezoelectric constants and field-induced strains – an order of magnitude larger than those of conventional piezoelectric ceramics – have been reported for '*domain-engineered*' (001)-oriented PMN-x%PT and PZN-x%PT crystals for compositions close to a morphotropic phase boundary (MPB). The MPB is supposed to be a near-vertical boundary in the x-temperature (x-T) field, separating rhombohedral (R) and tetragonal (T) phases. For example, in (001)-oriented PMN-x%PT crystals, the composition x=0.33 lies at the MPB, possess the optimum piezoelectric ($d_{33}$~2500 pC/N) and electromechanical coupling ($k_{33}$~94%)[3] coefficients.

Understanding of the structural origin of the high electromechanical properties of MPB compositions has undergone an evolution in thought. Early investigations attributed the high electromechanical properties of Pb(Zr$_{1-x}$Ti$_x$)O$_3$ ceramics to domain (or '*extrinsic*') contributions[4]. More recently, coincidental with their discovery of the high electromechanical properties in oriented PMN-x%PT and PZN-x%PT crystals, Park and Shrout[1] conjectured that the ultrahigh strain under applied electric field (E) was due to a R→T phase transition induced by E; however, the slim-loop nature of the ε-E curves is not conventional of an induced transition that is generally expected to be strongly hysteretic.

---

*Permanent address: Institute of Material Research, Tohoku University, Sendai 980-8577, Japan.

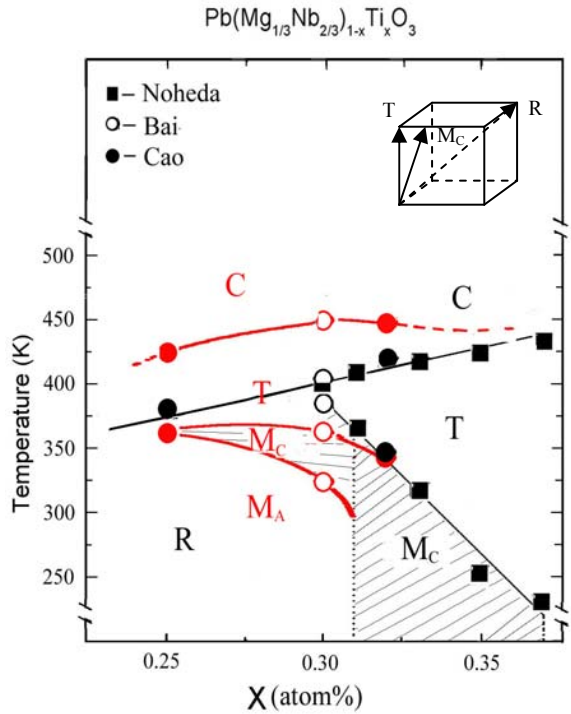

Figure 1 Modified phase diagram of PMN-xPT around the MPB according to recent data by Noheda et al.[14]. The solid line indicating the transition to cubic phase is reported by Noblanc et al., black solid squares by Noheda et al., open circles by Bai et al.[15], and solid circles by Cao et al. in this study. All black symbols and lines stand for E=0kV/cm and all red ones for E=2kV/cm. The polarization vectors of Monoclinic C ($M_C$), Tetragonal (T) and rhombohedral (R) in the perovskite unit cell are shown as an inset.

Subsequently, x-ray (XRD) and neutron diffraction experiments have shown the existence of various monoclinic (M) bridging phases in $PbZr_{(1-x)}Ti_xO_3$ ceramics[5-7], and in oriented PZN-x%PT[8-12] and PMN-x%PT[12-15] crystals. Two monoclinic phases, $M_A$ and $M_C$, have since been reported in PZN-x%PT [8-12]. The $M_A$ and $M_C$ notation is adopted following Vanderbilt and Cohen[16]. Recent neutron diffraction studies of the effect of an electric field (E) on PZN-8%PT by Ohwada et al.[11] have shown that a cubic (C)→T→$M_C$ transformational sequence occurs when field-cooled (FC), and that an R→$M_A$→$M_C$→T sequence takes place with increasing E at 350K beginning from the zero-field-cooled (ZFC) condition.

Similar $M_A$ and $M_C$ phases have also been reported in PMN-x%PT[12-15]. A recent study by Bai et al.[15] established that PMN-30%PT has a C→T→$M_C$→$M_A$ sequence in the FC condition, and a R→$M_A$→$M_C$→T one with increasing E beginning from the ZFC. Optical domain studies also have shown the existence of M phase in PMN-33%PT crystal by Xu et al[17]. Figure 1 summarizes the modified phase diagram of PMN-x%PT in the FC condition; which is re-plotted according to recent data published by Noheda et al.[14] and by Bai et al.[15] (alongside that to be presented in this paper). All black symbols represent XRD data taken under E=0kV/cm and all red ones for E=2kV/cm. The polarization vectors of the $M_C$, T, and R phases within the perovskite unit cell are shown in the inset of this figure. Interestingly, an anomalous shift of $T_C$ towards higher temperatures under electric field (E) was previously reported for PMN-30%PT[15], as illustrated in this figure. However, it is not yet know if this shift occurs only in a limited phase field in which the transformational sequence in the FC condition is C→T→$M_C$→$M_A$, where there is limited ranges of T and $M_C$ phase stability; or whether, the increase of the C→T boundary with increasing E may be characteristic of a wider phase field, in which the T and $M_C$ phase stability are favored.

In this investigation, we have carefully performed XRD studies to characterize the structure of the composition PMN-32%PT that is located on the PMN-rich side of the MPB as a function of temperature under various electric fields. Here, we report that the shift of the C→T phase boundary to higher temperatures with increasing E is a common characteristic of crystals whose compositions are in the vicinity of the MPB, irrespective of the width of the T and $M_C$ phase regions.

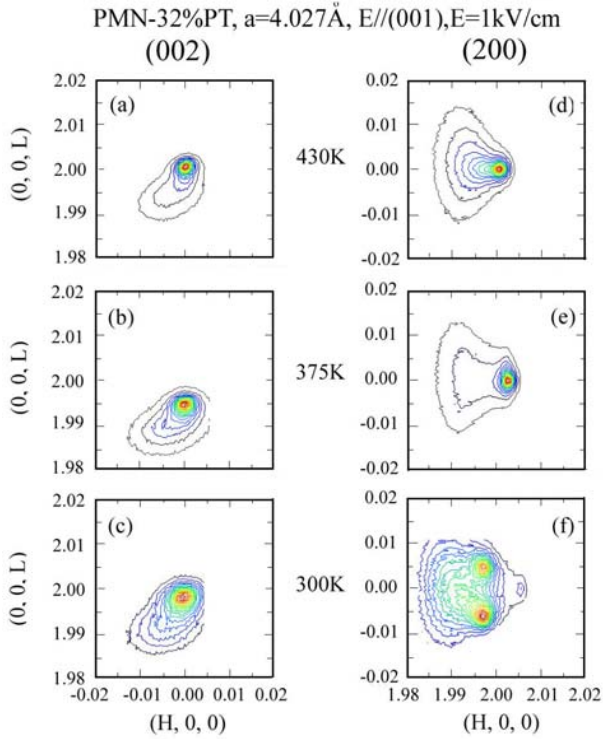

Figure 2  Mesh scans around the (002) and (200) profiles for PMN-32%PT at 430K, 375K, and 300K under as E=1kV/cm on cooling.

Crystals of PMN-32%PT with dimension of 3×3×3 mm³ were obtained from HC Materials (Urbana, IL), and were grown by a top-seeded modified Bridgman method. All surfaces were oriented along (100) pesudocubic faces, and were polished to 0.25μm. Gold electrodes were deposited on one pair of opposite surfaces of the cube by sputtering – we designate here the electroded faces as (001). Dielectric measurements were performed using a multi-frequency LCR meter (HP 4284A) to assure that the Curie temperature ($T_C$) of samples were close to that shown the phase diagram given by Noheda et al.[14]. XRD studies were performed using a Philips MPD high-resolution system equipped with a two bounce hybrid monochromator, an open 3-circle Eulerian cradle, and a doomed hostage. A Ge (220)-cut crystal was used as an analyzer, which had a θ-resolution of 0.0068°. The x-ray wavelength was that of $Cu_{K\alpha}$ = 1.5406Å and the x-ray generator were operated at 45kV and 40mA. The penetration depth in the samples was on an order of 10 microns. Each measurement cycle was begun by heating up to 550K to depole the crystal, with measurements taken on decreasing temperature. Measurements made under zero-field-cooling are designated as ZFC, whereas those made under field-cooling are designated as FC. At 450K, the lattice constant of PMN-32%PT was a=4.027Å, correspondingly the reciprocal lattice unit (or 1 rlu) was $a^*=2\pi/a=1.560Å^{-1}$. All mesh scans of PMN-32%PT shown in this study were plotted in reference to this reciprocal unit.

Figure 2 shows mesh scans taken on cooling under E=1kV/cm. Scans taken at 430K, which is close to the C→T transition, revealed somewhat broadened contours. However, the lattice parameters extracted from these (002) and (200) scans were nearly identical. Possibly, over a narrow temperature range near $T_C$, there is a small degree of C and T phase coexistence. With decrease of temperature, the (002) peak shifted towards lower L values, and the (200) peak towards higher H values.

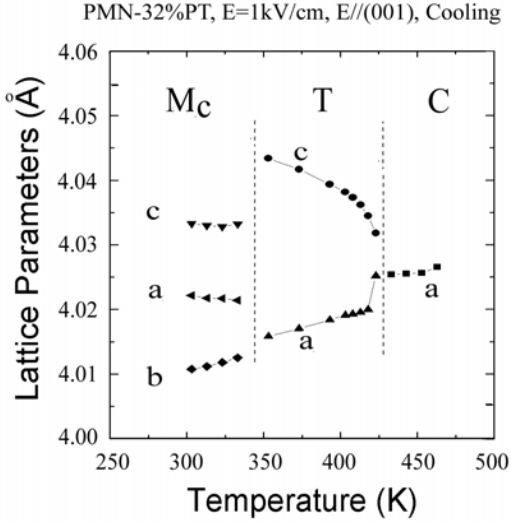

Figure 3  Evolution of lattice parameters as a function of temperature for PMN-32%PT under E=1kV/cm on cooling.

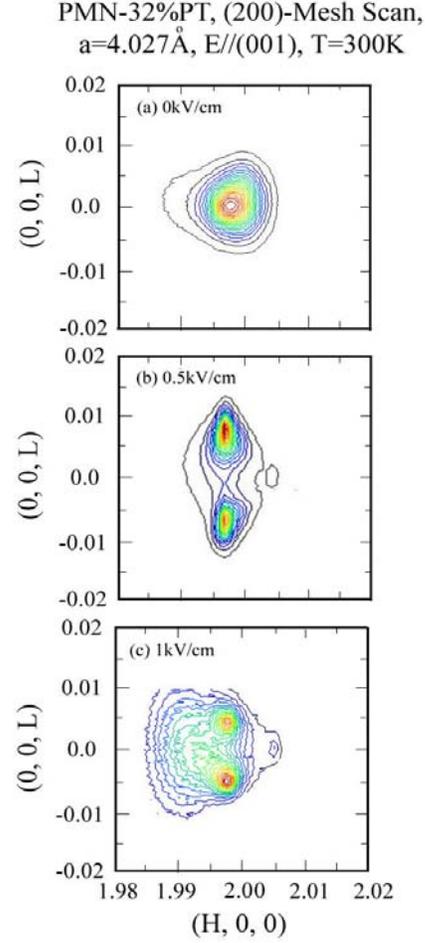

Figure 4 Electric-field dependence of the (H0L) contour around (200) obtained at 300K.

This demonstrates a C→T transition on cooling. However, as shown in the (200) mesh scan at 375K, an additional weak peak is present, indicating some 90°-domain formation along the <101>. This is possibly due to the limited penetration depth of our x-ray probe, but either way we did not observe a fully-aligned single domain configuration. Upon further cooling to 300K, the (200) reflection was found to split into three peaks – two (200) peaks, and a single (020) one; whereas, the (002) reflection remained as a single peak. Clearly, the (200) and (002) mesh scans at 300K have the signature features of the $M_C$ phase. The lattice parameters as a function of temperature on cooling under E=1kV/cm are plotted in Figure 3. At 430K, a decrease in the $a$-parameter was found at the C→T transition. Near the T→$M_C$ transition at ~350K, $c_M$ decreased with respect to $c_T$, $a_M$ increased with respect to $a_T$, and $b_M$ was nearly equal to $a_T$. In general, we found the temperature dependent lattice parameters for PMN-32%PT cooled under E=1kV/cm to be nearly identical to corresponding ones for ceramics in the ZFC condition[14] – both exhibited stable $M_C$ phases at 300K, with similar values of the lattice parameters. Figure 4 shows the evolution of (200)-mesh scans of PMN-32%PT with increasing electric field, taken at 300K in the FC condition. The (200)-mesh scan taken under zero fields is given in Figure 4(a). A single contour can be seen that is quite broad, possessing a rather long tail that extends along the longitudinal direction. In addition (data not shown), we did not observe any splitting in the (220)-mesh scan, and thus can rule out the possibility that the R phase is stable. Figures 4(b) and (c) show the (200)-mesh scan under different electric fields. Under a small field of E=0.5kV/cm, the (200)-mesh scan exhibited the signature pattern of the $M_C$ phase; and upon increasing the field to E=1kV/cm, the monoclinic pattern became more pronounced. According to the ZFC phase diagram reported by Noheda et al.[14] and redrawn in Figure 1, PMN-32%PT is located inside of the MPB region, where the monoclinic $M_C$ phase is present at T=300K over the compositional range of 31≤x≤37at%. Our results at ZFC did not show the $M_C$ type splitting, very likely because one or more of the $M_C$ domains were missing, which is quite common in single crystal diffraction measurements. The signature of $M_C$ phase becomes clear with FC, where the field helps stabilize different $M_C$ domains. We thus infer that the phase transformational sequence is C→T→$M_C$ in both the ZFC and FC conditions. Figure 5(a) shows the dielectric constant as a function of temperature for 350<T<460K was taken under different electric fields in the FC condition. This data was taken on cooling using a measurement frequency of 1 kHz. A single transition can clearly be seen in ZFC condition near 410K. According to the ZFC phase diagram[14] (redrawn in Fig. 1), this transition is the C→T one. Unlike prior results for PMN-30%PT[15], our dielectric peaks for PMN-32%PT were relative sharp near $T_C$, and only weakly frequency dependent (data not shown). In this regard, PMN-32%PT in the ZFC condition exhibits transition characteristics similar to those of a normal ferroelectric, rather than those of a relaxor. However, the C→T phase transition temperature, as determined by field dependent dielectric constant measurements, was *not* altered with increasing E. Although, in the FC condition, the magnitude of the dielectric constant was dramatically decreased in the T-phase region by field-cooling, even under a small field of E=0.5kV/cm.

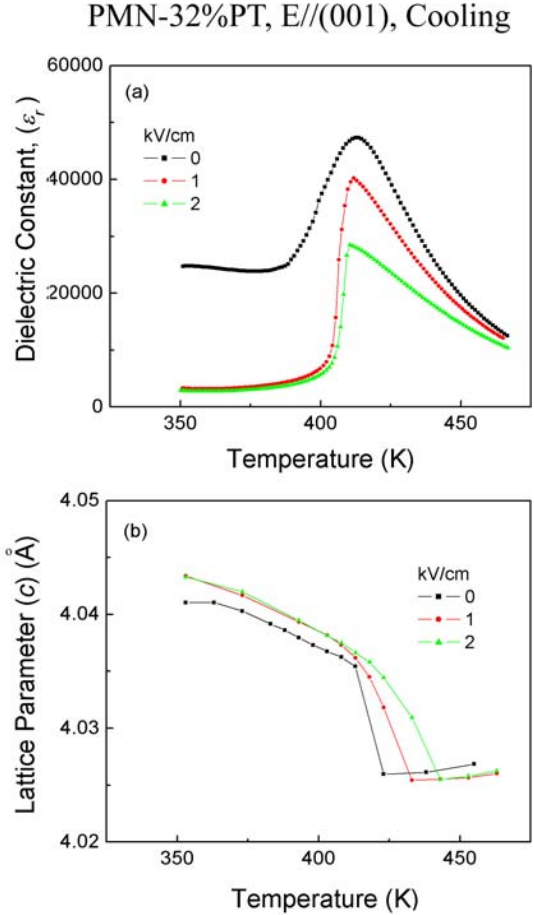

Figure 5 Temperature dependence of (a) the dielectric constants at 1 KHz and (b) lattice $c$-parameter derived from (002) reflection under different levels of electric fields.

Figure 5(b) shows the evolution of the lattice $c$-parameter as a function of temperature at different electric fields. Here, we defined $T_C$ as the temperature at which the lattice constant $c$ begins to increase in magnitude upon cooling. In this figure, it can clearly be seen that the C→T transition shifts towards higher temperature with increasing E. We determined the rate of increase in $T_C$ for PMN-32%PT to be ~10K·cm/kV In addition, no abnormal changes in the lattice parameter values or its slope can be seen in Fig. 5(b) for temperatures below 410K. This indicates that there is a region where the T and C phases coexist, which is also consistent with our above observations in Figure 2 concerning the mesh scans at 430K.

An important observation from this work for PMN-32%PT is an apparent difference between $T_C$ as determined by comparisons of dielectric and structural measurements in the FC condition. We summarize in the PMN-x%PT phase diagram of Figure 1 the shift in the C→T boundary upon application of E=2kV/cm (shown in *red*), as determined by XRD. The corresponding C→T boundary, determined from dielectric measurements taken under E=2kV/cm, follow that of the ZFC condition (shown in *black*). Also summarized in this figure is the relative magnitude of the shift in the C→T phase boundary over a wide composition field (note: all XRD data). It is relevant to notice that the C→T boundary shift-rate was reduced as the MPB was approached, with increasing x. For example, we observed a shift-rate of $\delta T_c/\delta E$~10 K·cm/kV for PMN-32%PT; whereas, the shift-rate for PMN-30%PT was previously reported to be ~25 K·cm/kV[15]. Furthermore, in this investigation, we found PMN-25%PT (which has relaxor characteristics) to have an identical rate to that for PMN-30%PT with $\delta T_c/\delta E$~25 K·cm/kV (data not shown, but summarized in the phase diagram). Our summary of results demonstrates that the increase of the C→T boundary with increasing E is not limited to a phase field with narrow ranges of T and $M_C$ stability; but, rather is seemingly a common characteristic of crystals whose compositions are in the vicinity of the left-hand side of the MPB, irrespective of the width of the T and $M_c$ phase regions. Although, the value of $\delta T_c/\delta E$ is reduced as one approaches the MPB, and crosses over into the T-phase region on the right-hand side of the MPB.

One possible explanation for the dependence of the C→T phase boundary on E is that polar nano-regions (PNR) exists near $T_C$, for compositions on the left-hand side of the MPB. Application of E along (001) might then readily favor an alignment of PNR whose polarization is oriented along the c-axis. The observed shift in the C→T phase boundary could then simply reflect a change in the relative population of tetragonal PNR variants under E. Due to the diffuse nature of the transition, the volume fraction of tetragonal PNR would gradually increase on cooling over a relatively broad temperature range, allowing for gradual lattice parameter changes. However, for PMN-32%PT whose c/a ratio is larger in the T phase than PMN-30%PT, the coexistence of tetragonal domains with the cubic phase would be suppressed by a significantly higher elastic energy density, i.e., $\sim(c/a)^2$. Thus, the C→T phase transformation near and above the MPB would be sharper, and its phase boundary more difficult to shift under E.

In summary, structural and dielectric measurements of (001)-oriented PMN-32%PT crystals have been performed. A C→T→$M_C$ sequence was found in both the ZFC and FC conditions. However, an important change was observed in the structural data with increasing E – an anomalous increase in the C→T boundary with increasing E was found, which becomes less pronounced on approach the MPB.

We would like to gratefully acknowledge financial support from the U.S. Department of Energy under contract No. DE.-AC02-98CH10886 and the Office of Naval Research under grants N000140210340, N000140210126 and MURI N0000140110761. We would also like to thank HC Materias for providing the single crystals used in this study.